\documentclass[twocolumn,showpacs]{revtex4}	% for arXiv submission
\usepackage{graphicx}

\begin{document}

%\title{Discrete optical hyper-solitons in photonic lattices}
\title{Lattice hyper-solitons in photorrefractive materials}
\author{Jos\'e Ram\'on Salgueiro, Humberto Michinel and Mar\'{\i}a I. Rodas-Verde}
\affiliation{\'Area de \'Optica, Facultade de Ciencias de Ourense,\\
Universidade de Vigo, As Lagoas s/n, Ourense ES-32004, Spain}

\begin{abstract}
We show a novel kind of nonlinear waves in two-dimensional photonic lattices. 
This waves take the form of light clusters that may fill an arbitrary number of
lattice sites. We have demonstrated by numerical simulations that stable 
propagation can be achieved under adequate conditions and we have described the 
unstable patterns developed otherwise. Our results show that these new kind of 
nonlinear waves can be easily found in current experiments.

\end{abstract}

%\ocis{190.5530,190.5330}
%\pacs{}

\maketitle

Optical lattices are artificial crystals of light\cite{chr_nature}
that can be obtained by shining an adequate optical material (usually 
photorrefractive) with a number of mutually coherent laser beams. The 
interference pattern created by those beams, induces a 
spatially periodic modulation of the refractive index in the material. This 
effect can be used to trap a probe laser beam, usually much weaker and of 
different frequency than those beams used to generate the lattice. 

The study of nonlinear waves in periodic potentials has become an active 
field of research in recent years due to its applications in control and 
manipulation of light properties. In this kind of systems, the appropriate 
combination of the probe beam power and the lattice depth 
and period yields to stable and localized light propagation for the probe 
laser. This beam, under adequate conditions, can generate a 
{\em lattice soliton}\cite{chr_ol}, linked to a site of the periodic potential, 
that propagates undistorted. 

The generation of lattice solitons has been demonstrated experimentally 
in both one-dimensional \cite{FleisPRL,NeshOL} and 
two-dimensional\cite{FleisNature,Martin}
potentials. Novel kinds of nonlinear waves like soliton dipoles and
quadrupoles\cite{chen_dipoles,siam_dipoles} as well as discrete 
vortices have been also demonstrated\cite{prl_our,prl_moti}.
Most of this results have been generalized for the case of several 
probe beams that are mutually incoherent, yielding to the so-called 
{\em discrete vector solitons}\cite{chen_vector}.

In this Letter, we will show that all these previous distributions belong
to a more general and novel kind of nonlinear waves that
exists in two-dimensional periodic lattices. This waves are clusters 
of parallel laser beams that are linked to the sites of the lattice. 
The adequate choice of beam powers and lattice parameters yields to the
formation of arrays of lattice solitons that propagate without shape distortion 
for arbitrarily long distances. The extension of this light distributions can be
arbitrary large, filling an unlimited number of lattice sites, so they may be 
regarded as hyper-solitons. 

We will calculate the explicit form of these distributions and demonstrate that 
stable propagation is achieved for a wide range of configurations. We will also 
test the stability of these structures when they are generated using arrays of 
Gaussian beams\cite{karta,rodas-verde} with adequate powers, widths and relative 
phases, calling in that way for the experimental demonstration.

We will study the paraxial propagation along $z$ of a laser beam with 
envelope $\Psi(x,y,z)$, through a  photorrefractive material with a periodic 
modulation of the refractive index in the transversal plane $(x,y)$. This process 
can be described by the following wave equation \cite{chen_vector},
\begin{equation}
i \frac{\partial \Psi}{\partial z} + \Delta_{\perp} \Psi-
     \frac{\Psi}{1+V(x,y)+|\Psi|^2}=0,
     \label{model}
\end{equation}
where $\Delta_{\perp}$ is the transverse Laplacian operator, the spatial 
variables $x$ and $y$ are expressed in units of the wavelength 
($\lambda$) and $z$ is measured in units of $4\pi\lambda$. 
The function $V(x,y)=A\cos^2(\pi x/T)\cos^2(\pi y/T)$ defines the 
periodic optically-induced lattice, where $A$ and $T$ are respectively its 
amplitude and period. Both parameters can be easily controlled by acting on 
the laser beams that generate the grid. 

\begin{figure}[tbp]
  \centerline{
    \scalebox{0.4}{
      \includegraphics{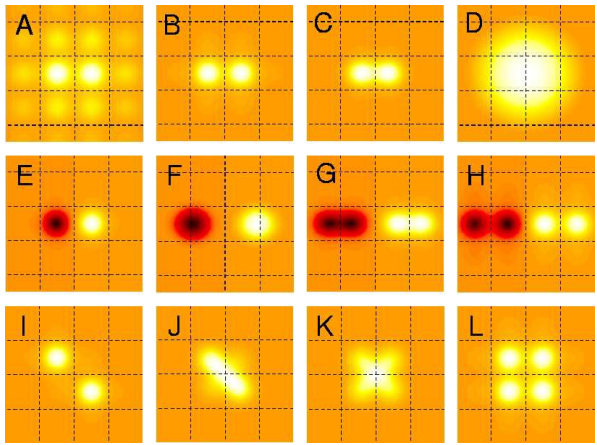}}
  }
\caption{Plots of the amplitude of different multi-hump solitons, 
illustrating the dependence of their shapes with power (or $\beta$). 
Dashed lines indicate the nodes of the lattice. Labels correspond 
to points in Fig.~\ref{pow}. }
\label{st1}
\end{figure}

We look for stationary solutions of the form $\Psi(x,y,z)=u(x,y)\exp(i\beta z)$, 
where $\beta$ is the propagation constant. These solutions can be numerically 
calculated and constitute families of solitons described by the parameter 
$\beta$. In Fig.~\ref{st1} we show some examples of stationary states for the 
particular values of the parameters, $A=1$, $T=10$. In order to have localized 
solutions with $u(x,y\rightarrow \pm \infty)=0$, due to the particular form of the 
potential term in Eq. (\ref{model}), it is required that $\beta<0$. 

\begin{figure}[tbp]
  \centerline{
    \scalebox{0.35}{
      \includegraphics{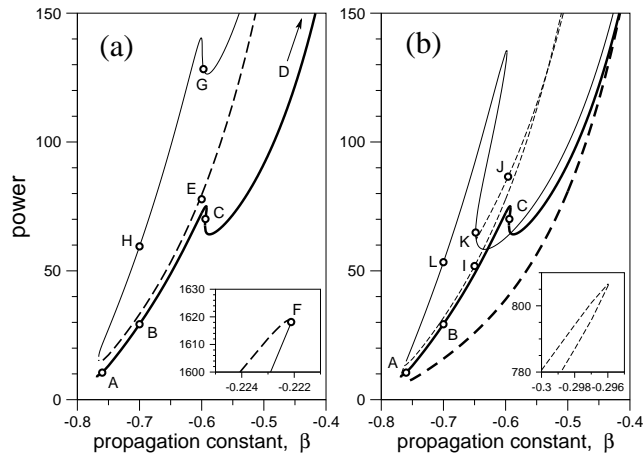}}
  }
\caption{Power curves for different families of multi-hump solitons. (a) Solitons 
with two side by side light spots (continuous thick line: two same-phase lobes; 
dashed line: same but inverted phase; continuous thin line: two double-hump 
spots with inverted phase). (b) Solitons with lobes of the same relative phase 
and disposed in different configurations (thick dashed: one lobe; thick 
continuous: two side by side lobes; thin dashed: two diagonally disposed lobes; 
thin continuous: four lobes disposed in a two by two array). Letters label points 
corresponding to the examples in Fig.~\ref{st1}. Point D, out of the frame, has 
coordinates $(-0.02,1.45\times 10^{5})$.}
\label{pow}
\end{figure}

Each family of solitons is topologically defined by the number of lobes and 
phase configuration. In Fig.~\ref{pow}, we show plots of the power 
$P=\int_{-\infty}^\infty|u|^2dxdy$ versus the propagation constant for several 
families of solitons, corresponding to different cases illustrated in 
Fig.~\ref{st1}.

In the linear limit, the periodicity of the medium requires Bloch-type periodic 
solutions (see Fig.~\ref{st1}-A), which extend over the whole lattice. If the 
nonlinear effect is significant, localized solutions appear, taking the form of 
multi-hump solitons whose lobes are located at the maxima of the lattice. It can 
be appreciated that families of solitons with a higher number of lobes present a 
higher power. For a particular family, the shape and amplitude of the lobes 
depend on the value of the propagation constant, so that $P$ basically increases 
with $\beta$.


In Fig.~\ref{st1}-B,C it is shown that as power raises, two close lobes of 
the same sign undergo fusion into a single lobe state (Fig.~\ref{st1}-D) 
centered at a lattice nodal line, which may reach unlimited high values. This 
is due to the nonlinear effect which shades the lattice for high powers. 

Modes with two inverted-phase lobes (Fig.~\ref{st1}-E) exhibit a different 
behavior with power. Neighbor lobes prevent each other from growing up, 
increasing power arbitrarily with $\beta$. Instead, the lobes drift away from 
each other, also increasing their width, up to the point where they are located 
on lattice nodal lines (Fig.~\ref{st1}-F). There, the power curve reaches an end 
point where branches corresponding to other families join 
(see Fig.~\ref{pow}-(a)). This is the case for the family of solitons composed by 
four lobes disposed in a two double-lobe configuration (Fig.~\ref{st1}-H). 
For this family, increasing $\beta$ yields to a fusion of the same-sign 
doublets (Fig.~\ref{st1}-G) ending with the state in Fig.~\ref{st1}-F.    

Fusion and splitting of lobes always correspond to an inversion in the sign 
of the slope of the $P$ vs $\beta$ curve and this inverted-slope section of the 
curve is more pronounced as many lobes 
are involved in the fusion (see for instance thick and thin continuous lines in 
Fig~\ref{pow}-(b), corresponding to two and four lobe family respectively). 
Analogous fusion or splitting take place for states with 
diagonally-disposed 
lobes (see Fig.~\ref{st1}-I). In this case, the fusion takes place towards a 
fundamental soliton centered at a node in the common corner of both lattice 
sites (Figs.~\ref{st1}-J,K). 

\begin{figure}[htbp]
  \centerline{
    \scalebox{0.4}{
      \includegraphics{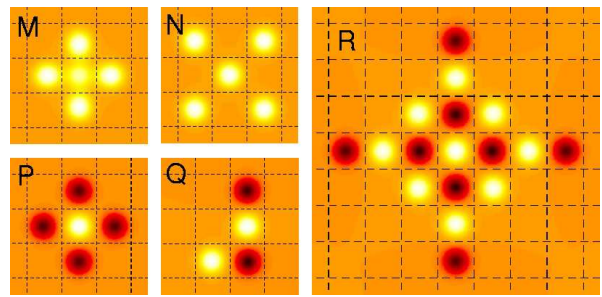}}
  }
\caption{Examples of higher-order solitons, showing different configurations, 
including one with a large number of humps (labeled R). For all of them 
$\beta=-0.5$. Dashed lines indicate the zeros of the lattice.}
\label{st2}
\end{figure}

In cases where an odd number of symmetrically disposed lobes does not 
permit a preferential fusion of two of them, a new scenario appears. 
For example the family of soliton M (Fig.~\ref{st2}), changes when $\beta$ is 
increased, lowering the amplitude of the central lobe up to the point where it 
disappears. Soliton N, however, does not qualitatively change. 

States with a higher number of lobes present more complicated dependences 
of the shape with power, and they can be basically accounted in terms of the 
same effects described above. An example is the two-by-two lobe soliton (L) 
whose lobes also merge together at the negative slope region of the curve. On 
the other hand, branches corresponding to different solitons may join together 
at points where fusion of lobes produce the same final soliton. This is the case 
at point (K) where the two and four lobe solitons fuse into a fundamental one.  

\begin{figure}[htbp]
  \centerline{
    \scalebox{0.35}{
      \includegraphics{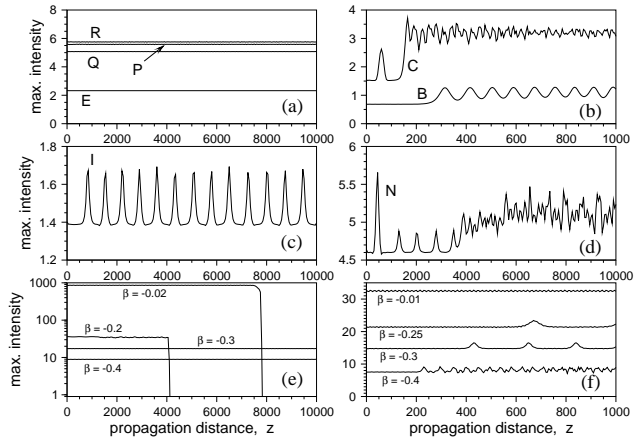}}
  }
\caption{Maximum amplitude versus propagation constant for a number of states. 
Line labels correspond to the solitons in Figs.~\ref{st1} and \ref{st2}.}
\label{propag}
\end{figure}

The stability of those hyper-solitons was investigated by means of 
numerical simulations using a standard Beam Propagation Method. The evolution 
of the maximum amplitude for some states is plotted in Fig.~\ref{propag}. The 
first conclusion is that those states whose lobes alternate sign are completely 
stable (for instance, all those corresponding to the dashed line in 
fig.~\ref{pow}(a)). Lines E, P, Q and R in Fig.~\ref{propag}-(a), correspondent 
to the same label solitons in Figs.~\ref{st1} and \ref{st2}, confirm the 
stability of those states. The numerical simulations were also carried out using 
arrays of Gaussian beams instead of the exact eigenstates and they resulted 
stable as well. This robustness would eventually allow to reproduce the results 
experimentally and make this beams suitable to be used for optical control 
operations.    

Contrariwise, solitons with lobes of the same sign or different sign pattern 
develop a number of instability scenarios which can be summarized 
as follows. For low power, the beam lobes are centered at lattice sites and 
they maintain their position developing power oscillations due to the coupling 
taking place from one to each other (see line B in Fig.~\ref{propag}-(b); 
solitons H, L and M present a similar behavior). 
For states corresponding to the negative slope region of the power curve, 
close lobes merge together and oscillate laterally with an amplitude of the 
order of the lattice pitch. This transversal oscillations also affect the 
amplitude (line C in Fig.~\ref{propag}-(b); similar behavior is found for 
soliton G). Similar oscillations are also found for solitons in the 
region of the power curve where the slope turns again positive, for moderate 
power. For high power, however, the lobes are found to fly off from each other 
after some propagation distance.

A different scenario is that of states with lobes disposed diagonally. 
Alternating sign ones remain stable. For those with same sign lobes, however, 
the instability pattern is different according to the branch of the power curve. 
Those belonging to one of the branches (solitons I and N for instance), develop a 
periodic phase (sign) inversion between alternate lobes (see line I in 
Fig.~\ref{propag}-(c)). Nevertheless, solitons of higher order develop for longer 
propagation distance more complicated coupling patterns (see line N in 
Fig.~\ref{propag}-(d)). On the other hand, states belonging to the other branch 
(like soliton J) develop a fusion of lobes followed by transversal oscillations, 
similarly to the case of negative slope.     
   
Concerning the fundamental soliton (one lobe) it is stable for low powers when 
centered at a lattice maximum (see simulations for different powers in 
Fig.~\ref{propag}-(e)). For higher powers, however, it wanders around, since the 
lattice is not enough to clamp it to the site, flying away at the end. The 
propagation distance at which the lobe starts to move increases as power 
increases and the soliton becomes virtually stable as power tends to 
infinite. A similar behavior is found when the state is centered on a nodal 
line (soliton D for instance) or nodal point, but in this case it is unstable 
for low powers, developing lateral oscillations between two nodal positions (see 
Fig.~\ref{propag}-(f), simulations for different power D-type solitons).

% --------------------------------------

In conclusion, we have introduced and analyzed novel types of stable discrete 
solitons in two-dimensional photonic lattices. Our simulations reveal that the 
experimental demonstration of our results is accessible with standard techniques.

The authors acknowledge support from the Ministerio de Educaci\'on y Ciencia of 
Spain (projects FIS2004-02466, FIS2004-20188-E and Ram\'on y Cajal contract 
granted to J. R. Salgueiro) and from Xunta de Galicia (project 
PGIDIT04TIC383001PR).

\end{document}